\documentclass[traditabstract]{aa} % for the abstract without structuration 
\usepackage{graphicx}
\usepackage{subfig}
\usepackage{txfonts}
\usepackage{natbib}
\newcommand{\Hi}{\textup{H\,{\mdseries\textsc{i}}}}
\newcommand{\HI}{\textup{H\,{\mdseries\textsc{i}}}~}

\newcommand{\Msun}{{M$_{\odot}$}}

\def\kms{{km~s$^{-1}$}}

\def\degr{\hbox{$^\circ$}}

\def\arcsec{\hbox{$^{\prime\prime}$}}
\def\fdegr{\hbox{$.\!\!^\circ$}}
\def\farcmin{\hbox{$.\mkern-4mu^\prime$}}
\def\farcsec{\hbox{$.\!\!^{\prime\prime}$}}
\begin{document}
   \title{Centaurus~A: morphology and kinematics of the atomic hydrogen}

   \subtitle{}

   \author{C. Struve\thanks{e-mail: struve@astron.nl}
          \inst{1,2},
          T.A. Oosterloo\inst{1,2},
          R. Morganti\inst{1,2}
	  \and
          L. Saripalli\inst{3,4}
          }

%   \offprints{C. Struve}

   \institute{Netherlands Institute for Radio Astronomy,
              Postbus 2, 7990~AA, Dwingeloo, The Netherlands
         \and
             Kapteyn Institute, University of Groningen,
             Landleven 12, 9747~AD, Groningen, The Netherlands
	 \and
	     Raman Research Institute, CV Raman Avenue,
             Sadashivanagar, Bangalore 560080, India
	 \and
             CSIRO, Australia Telescope National Facility,
             PO Box 76, Epping NSW 1710, Australia
             }

   \date{Received 04 March 2010; accepted 08 March 2010}

  \abstract {We present new ATCA 21-cm line observations of the neutral hydrogen in the nearby radio galaxy Centaurus~A. We image in detail (with a resolution down to 7\arcsec , $\sim 100$ pc) the distribution of \HI along the dust lane. Our data have better velocity resolution and better sensitivity than previous observations. The \HI extends for a total of $\sim 15$ kpc. The data, combined with a titled-ring model of the disk, allow to conclude that the kinematics of the \HI is that of a regularly rotating, highly warped structure down to the nuclear scale. The parameters (in particular the inclination) of our model are somewhat different from some of the previously proposed models but consistent with what was recently derived from stellar light in a central ring. The model nicely describes also the morphology of the dust lane as observed with Spitzer. 
There are no indications that {\sl large-scale} anomalies in the kinematics exist that could be related to supplying material for the AGN. Large-scale radial motions do exist, but these are only present at larger radii ($r > 6$ kpc). This unsettled gas is mainly part of a tail/arm like structure. The relatively regular kinematics of the gas in this structure suggests that it is in the process of settling down into the main disk. The presence of this structure further supports the merger/interaction origin of the \HI in Cen~A. From the structure and kinematics we estimate a timescale of $1.6 - 3.2 \times 10^8$~yr since the merging event.
No bar structure is needed to describe the kinematics of the \Hi . The comparison of the timescale derived from the large-scale \HI structure and those of the radio structure together with the relative regularity of the \HI down to the sub-kpc regions does not suggest a one-to-one correspondence  between the merger and the phase of radio activity. 
Interestingly, the radial motions of the outer regions are such that the projected velocities are {\sl redshifted} compared to the regular orbits. This means that the blueshifted absorption discovered earlier and discussed in our previous paper cannot be caused by out-moving gas at large radius projected onto the centre. Therefore, the interpretation of the blueshifted absorption, together with at least a fraction of the redshifted nuclear absorption, as evidence for a regular inner disk, still holds.
Finally, we also report the discovery of two unresolved clouds detected at 5.2 and 11~kpc away (in projection) from the \HI disk. They are likely an other example of left-over of the merger that brought the \HI gas.}

   \keywords{galaxies: active --
             galaxies: elliptical --
             galaxies: individual (Centaurus~A) --
             galaxies: kinematics and dynamics --
             galaxies: structure --
             galaxies: ISM
               }

   \authorrunning{C. Struve et al.}

   \maketitle
%
%________________________________________________________________

\section{Introduction}

The assembly of early-type galaxies and its connection to the presence of an active nucleus (and in particular a radio-loud nucleus) is still a matter of debate. Gas-rich galaxy mergers and interactions could play an important role in providing the fuel to trigger nuclear activity (including the radio phase). However, recent studies of radio galaxies have shown that the activity in some of these galaxies may be associated instead with the {\sl slow}  accretion of (hot) gas. This is thought to be the case in particular, for edge-darkened radio galaxies where advective low efficiency/rate flows could be the dominant mode of accretion \citep[see e.g.][]{allen06,balmaverde08}.

One way to address the issue of the relation between interaction/merger and AGN, is by studying  the gas in radio galaxies. Nearby sources are prime targets as the morphology and kinematics of the gas --- both at large scales and in the nuclear region --- can be studied in detail and related to each other. Atomic hydrogen can be a useful tracer for these purposes. In particular, \HI gas found at kiloparsec scales, or larger, can be used to trace the evolution of the host galaxies (e.g.\ major merger vs.\ small accretions). On the other hand, \HI in absorption is one of the diagnostics to probe the central regions of active galactic nuclei and the interaction of the AGN with its immediate environment. \HI absorption studies offer the unique possibility of tracing the nature of the accretion or the energetics of the outflow \citep[e.g.][]{morganti05}. 

Unfortunately, the number of radio-loud galaxies where the atomic neutral hydrogen can be studied in such details both in emission {\sl and} in absorption is limited. 
Due to its proximity\footnote{At the assumed distance of Cen~A of 3.8~Mpc \citep{harris10}, 1~arcmin corresponds to 1.1~kpc.}, the nearby radio galaxy Centaurus~A (Cen~A) is a unique object where to study the different phases of the gas, including \Hi , in a radio-loud AGN. This can be done on spatial scales ranging from the inner kpc, where the influence of the AGN can still be relevant, up to many kpc where the signature of past merger(s) might be still  visible through unsettled gas.

Cen A harbours a well known  radio source  classified as a Fanaroff-Riley type-I source \citep[FR-I,][]{fanaroff74} with a total power $P_{2.7\rm{GHz}}=10^{24.26}$~W~Hz$^{-1}$. The total extent of the radio source reaches more than 8\degr , making Cen A the largest extragalactic radio source in the sky \citep{cooper65}.
The radio source shows a variety of features on different scales. A radio jet extends from sub-parsec scales \citep[e.g.][]{jones96,horiuchi06} to a projected distance of $\sim 6$ kpc from the nucleus, ending in an inner radio lobe \citep{burns83,clarke92}. Vast regions of fainter emission extend beyond the inner lobes: the middle lobe extends out to 40 kpc, while the outer lobe is even more diffuse extending out to more than 500 kpc \citep[see Fig.~1 in][]{morganti99}. The different orientation and morphologies of the various structures could be the result of precession \citep[e.g.][]{haynes83} of the central engine combined with strong interaction with the surrounding medium. Alternatively, the nuclear activity in Cen~A has been recurrent, although recent spectral index studies suggest that the large-scale structure is currently undergoing (or has very recently undergone) some particle injection event \citep[][and refs. therein]{hardcastle09}. 

Most of the cold and warm gas in Cen~A is concentrated along the strongly warped dust lane \citep[e.g.][]{nicholson92,quillen06A}. The neutral hydrogen present in this region (both in emission and absorption) has been studied by e.g. \citet{vdh83,vangorkom90,sarma02}. The morphology and kinematics of the outer regions of the \HI disk suggest that some of the gas has not yet settled into regular rotating orbits \citep{vangorkom90}.

\HI has also been found well outside this disk. At a distance between 10 and 15 kpc from the nucleus, \citet{schiminovich94} have found a partial ring structure with a smooth N-S velocity gradient and rotating perpendicular to the gas located along the dust lane (but co-rotating with the stellar body). In the north-eastern region of this ring, young stars have been detected that have possibly formed by the interaction between the radio jet and the neutral hydrogen in the ring \citep{oosterloo05}.

Cen~A was shaped through hierarchical merging over many Gyr as revealed by the spread in ages of globular clusters \citep[see e.g.][]{woodley10}. The last major merger event happened about 5~Gyr ago \citep{peng04}.
The prominent warped dust and gas disk in the central region, and the outer partial ring structure, suggest that Cen~A experienced a recent merging event where its large elliptical body merged with a smaller gas-rich galaxy \citep{baade54, tubbs80}. \citet{quillen93} calculated a timescale of about 200 million years since the core of the infalling galaxy merged with the centre of the elliptical galaxy. This is consistent with the timescales derived by the presence of tidal debris and by the shell-like features containing atomic hydrogen \citep{schiminovich94,peng02} and molecular gas \citep{charmandaris00}.

%_____________________________________________________________
%                                             Two column Table
%_____________________________________________________________
%
\begin{table*}
\caption{Summary of observations.}
\label{reducprop}
\centering
\begin{tabular}{c l l}
\hline\hline
Total on-source integration time &  & $3 \times 12$~h\\
Observation dates &  & 12/14/22 April 2005\\
Bandwidth &  & 16~MHz\\
Number of channels in the cubes &  & 512\\
Channel spacing ($dV$) &  & 6.6~\kms\\
Velocity resolution ($\Delta v$, after Hanning) &  & 13.2~\kms\\
Shortest/longest baseline &  & 77~m/6~km\\
Bandpass/flux calibrator &  & PKS~1934--638\\
Gain/phase calibrator &  & PKS~1315--46\\
\hline
Data cube & high resolution & low resolution\\
Weighting scheme & uniform & robust 0\\
Beam (HPBW) & $8.1\arcsec\times 6.8$\arcsec & $19.6\arcsec\times 18.2$\arcsec \\
Beam position angle & --12.3$^{\circ}$ & --48.3$^{\circ}$\\
rms noise (mJy~beam$^{-1}$) in the cubes & 1.3 & 1.2\\
rms noise ($10^{19}$~atoms~cm$^{-2}$) & 30.92 & 4.45\\
rms noise ($M_{\odot}$~pc~$^{-2}$) & 2.48 & 0.36\\
Peak flux continuum (Jy~beam$^{-1}$) & 4.41 & 4.95\\
rms noise (mJy~beam$^{-1}$) continuum & 32.2 & 26.7\\
\hline
\end{tabular}
\end{table*}
%
%__________________________________________________________________
%
%__________________________________________________________________
%
\begin{figure*}
\centering
\includegraphics[width=0.7\textwidth, angle=270]{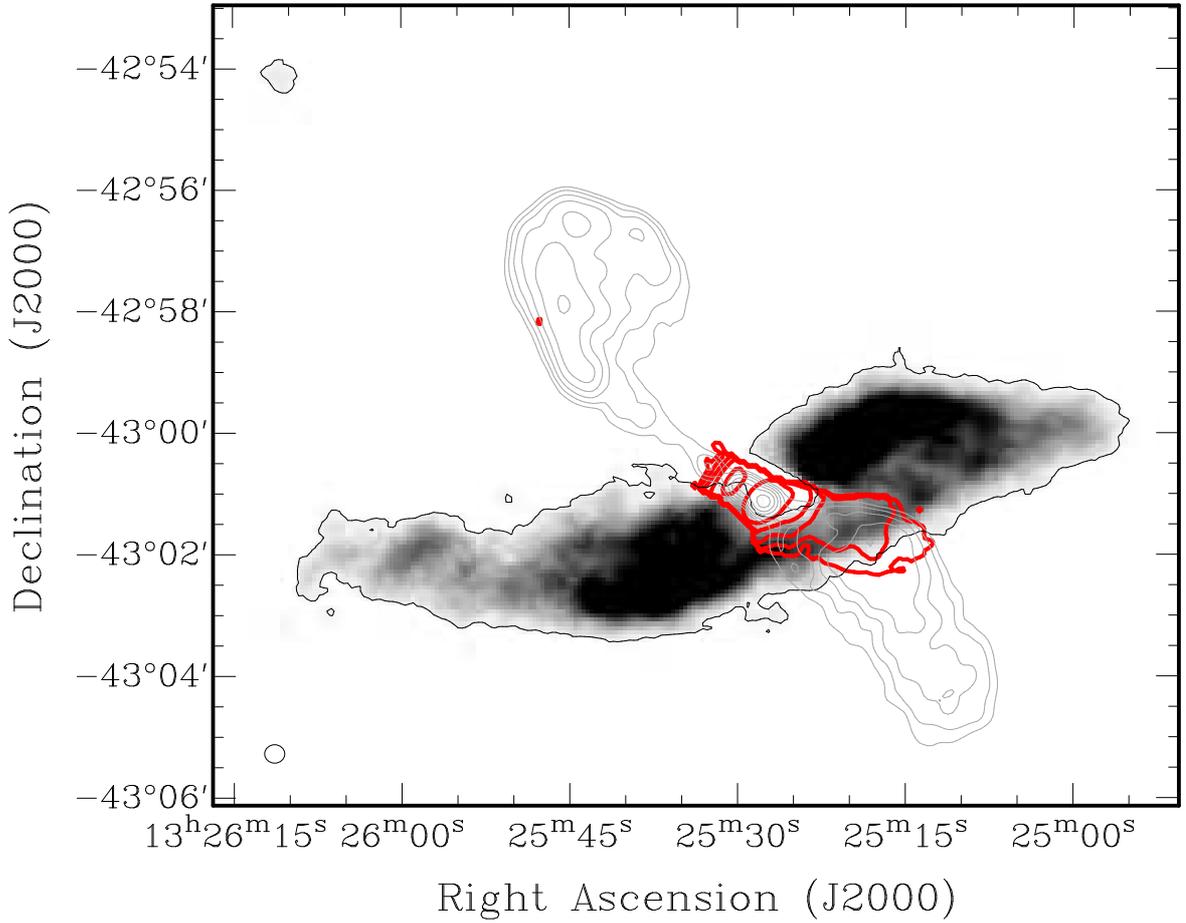}
\caption{\HI emission in greyscale overlaid with the low resolution radio continuum (thin, grey contour levels: 0.5~Jy~beam$^{-1}$ increasing with a factor of 1.5) and absorption contour levels on top (thick contours, red in the online version). Absorption contour levels: $0.33\times 10^{19}$~cm$^{-2}$ increasing with a factor 3. Emission contour level: $1.5\times10^{20}$~cm$^{-2}$.}
\label{m0.optical}
\end{figure*}
%
%__________________________________________________________________
%

\HI absorption is detected against the nucleus, the northern jet and the southern lobe \citep[e.g.][]{roberts70, vdh83,vangorkom90, sarma02}. The absorption against the jet and southern lobe has been interpreted as gas that is part of the large-scale disk which is roughly perpendicular to the jet axis \citep[$PA=50$\degr ,][]{tingay98}. 
Since no absorption is found against the northern radio lobe, this lobe is likely pointing towards us whereas the southern radio lobe points away. Van Gorkom et al. (1990) and \citet{burns83} concluded that the northern part of the gas disk is (at least partly) behind the radio source and the southern part of the \HI disk is in front.
For the component of \HI absorption against the nucleus, earlier observations have shown the presence of redshifted absorption against the (unresolved) core \citep[e.g.][]{vdh83,vangorkom90,sarma02}, but our recent results, discussed in \citet[][Paper~I]{morganti08} and in this paper, have shown that blueshifted absorption against the core is also present.

Despite the many studies devoted to Cen~A, there are  a number of open questions  that are essential for the understanding of the formation and evolution of Cen~A and its AGN. 
For example, the relation between the age of the radio structures and the timescale of the merger is not completely clear. 
In particular, it is unclear what triggers the current AGN activity (i.e. the activity producing the inner radio lobe) and whether it is related to the merging event. 
Thus, it is important to obtain the morphology and kinematics of the gas disk to the largest possible radii and investigate the structure of the disk. This can provide an estimate of the timescale of the merger and allow to investigate the presence and kinematics of unsettled gas (e.g. whether significant radial motions are present and, if so, on which scale).  At the same time, it is important to investigate the distribution of \HI down to the most inner parts  of Cen~A to understand whether kinematical signatures of gas connected with the fuelling of the AGN are present.

The large amount of \HI present in Cen~A and its proximity allow to obtain deeper observations which can be used to investigate the questions above. To achieve this, we have performed new spectral-line observations of Cen~A that have higher spatial and velocity resolution and better sensitivity than those of previous studies. These new observations have already allowed us to detect blueshifted absorption against the nucleus while also the redshifted absorption extends to higher velocities compared to previously known. These results have been presented in Paper~I. This gas has been interpreted as part of a circumnuclear disk, likely the counterpart of what was already detected in CO and other lines, leaving still open the question of what triggers the current AGN activity. To  further investigate this issue, we present in this paper the study of the kinematical state of the \HI disk and the results from 3D modelling.

The paper is organised in the following way. In Sect.~2 we present the new observations while in Sect.~3 we describe the results. In Sect.~4 we derive and discuss tilted-ring models of the \HI disk that are based on the 3D data cube that describe the strongly warped \HI disk as well as previously published data. We show that the \HI is dominated by rotation, with radial motions giving a significant contribution in the outer parts. Furthermore, we argue that the blue- and redshifted absorption against the nucleus must be mainly caused by gas close to the nucleus. In Sect.~5 we discuss the consequences of the observational and modelling results for the formation and evolution process of Cen~A, the merger timescales involved and the implications for the AGN activity. A short summary is given in Sect.~6.

%__________________________________________________________________ %
\begin{figure*} 
 \centering
  \includegraphics[width=0.90\textwidth]{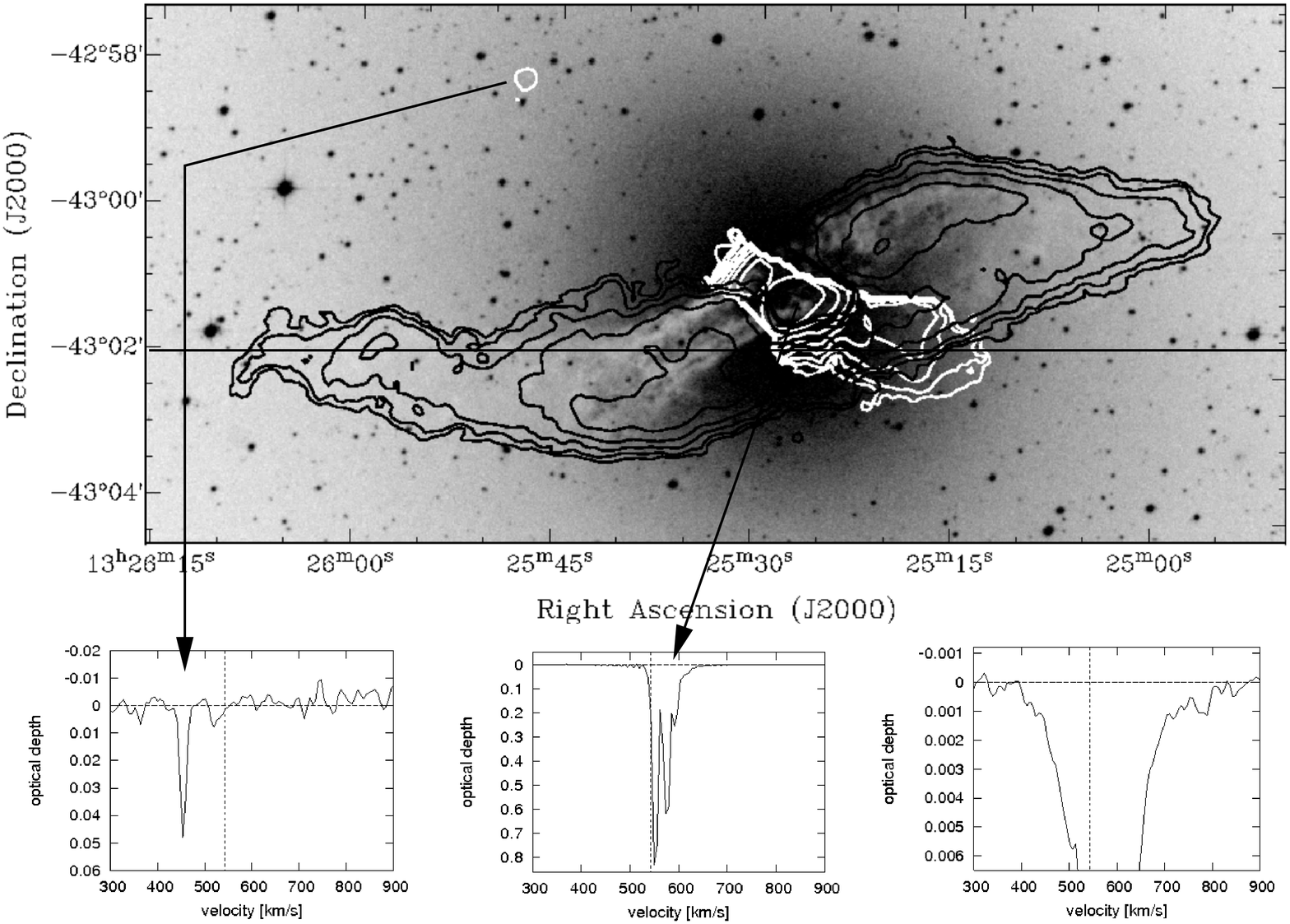} 
 \caption{Top panel: Total intensity contours over-plotted on DSS image. Contour levels: 0.2, 0.4, 0.8, 1.6 (absorption only), 3.2, 6.4, 12.8, 25.6 and $51.2\times 10^{20}$~atoms~cm$^{-2}$ for the emission (black) and absorption (white). $T_{\rm{spin}}=100$~K. The horizontal line indicates the position of the pv-slices in Fig.~\ref{pv.plot2} and \ref{pvplots.models}. Bottom left panel: spectrum of \HI absorption cloud. Bottom middle panel: spectrum of the nuclear absorption. Bottom right panel: zoom-in of bottom middle panel.}
 \label{mom0.spec} 
\end{figure*}
%__________________________________________________________________ 
%

\section{Observations and data reduction}

Cen~A was observed with the ATCA in April 2005. In Tab.~\ref{reducprop} some
basic information about the observations is summarised. These data were also presented in \citet{morganti08}. Three different standard configurations (750~m, 1.5~km and 6~km) were used, each  for a full 12-h run. This setup gives sufficient $uv$-coverage as well as  good spatial resolution. In addition, it avoids baselines shorter than 77~m. This is important because
the very strong, very extended continuum emission of Cen A makes bandpass
calibration in practice impossible on short baselines where the flux is well
over 100~Jy (see below).

The total bandwidth covered was 16~MHz - corresponding to a velocity range of
3000 \kms\ - with 512 frequency channels. This is much broader than used in
earlier observations \citep[e.g.][]{vdh83, vangorkom90, sarma02}. In order to
have enough line-free channels for the continuum subtraction, the central
velocity of the observing band was about 200 \kms\ below that of the systemic
velocity of Cen~A to account for the Galactic \HI emission/absorption. The
spectral resolution is 13.2~\kms\ after Hanning smoothing. We used PKS~1934--638
as bandpass and flux calibrator. To monitor the system gain and phase changes,
PKS~1315--46 was observed as a secondary calibrator for ten minutes every hour,
although self calibration was also employed (see below). The data reduction was
performed using the MIRIAD package \citep{sault95}.

The extremely strong radio continuum requires  careful bandpass calibration in
order to obtain  flat spectral baselines. Following a technique applied
successfully by \citet{oosterloo05} on earlier observations of Cen~A, we
smoothed the bandpass calibration obtained from PKS~1934--638 with a box-car
filter 15 channels wide. Since the instrumental features in the bandpass are
fairly broad in frequency, this procedure is allowed and it minimises the
contribution of extra-noise from the bandpass calibration that otherwise would
have occurred since the bandpass calibrator is fainter than the very extended
emission of Cen~A. Through this smoothing, the effective flux level of
PKS~1934--638 increases to about 60~Jy which is higher than the detected flux on
almost all baselines. The smoothed bandpass correction was applied to the
unsmoothed data of Cen~A. This procedure ensures that the velocity resolution is
preserved without increasing too much the noise of the data. The strongest \HI absorption is just over 2 Jy and the  dynamic range in the spectral line cubes is better than 1:1000.

The continuum was subtracted using UVLIN by making a second-order fit to the line-free channels of each visibility record and subtract this fit from the spectrum. The line cube, as well as the obtained continuum image, was cleaned. Hanning smoothing was applied to the line cube. Frequency independent self-calibration of the phase using the channel with the strongest nuclear absorption improved the image quality, but a few channels with the strongest absorption still show some calibration artifacts. The final cube was derived by combining the datasets of the three observations. The beam size in the resulting uniformly weighted cube is 8\farcsec1 $\times$ 6\farcsec8, PA~=~--12\fdegr3. The image noise is about 1.3~mJy~beam$^{-1}$. For the kinematic modelling of the disk (see Sect.~5), a cube with robust weighting set to 0 was made while also tapering the visibility data with a Gaussian taper. The resulting beam size for this cube is 19\farcsec6 $\times$ 18\farcsec2, PA~=~--48\fdegr3. This lower resolution cube is a compromise between spatial resolution and sensitivity.

\section{Results}

\subsection{The structure of the \HI emission}

As expected from previous observations, \HI is detected both in emission and in absorption and most of the gas  is concentrated in a relatively edge-on, warped  structure that mainly follows the dust lane (see Fig.~\ref{m0.optical} and \ref{mom0.spec}).  On both the eastern and western side, the \HI extends beyond the visible dust lane and, in total, has an extent of $\sim 15$~kpc in diameter. The total mass of the \HI disk  is $4.9\times 10^8$~\Msun . This is slightly higher than what was derived from VLA observations \citep[$4.2\times 10^8$~\Msun, after correcting for the difference in the assumed distance;][]{vangorkom90}. The difference is likely due to better separation of emission and absorption in the central regions (due to our better spatial resolution), and to the higher sensitivity of our observations.

In Sect.~4 we  present detailed modelling of the geometry and the kinematics of the \Hi , but we briefly discuss the main points here. The main conclusion is that the kinematics of the \HI is regular in the inner regions of the disk.
Large-scale radial motions exist, but  are only present at larger radii. There are no indications
of  anomalies  in the kinematics of the gas (on the scale down to the inner beam, $\sim 100$ pc)  that could be related
to supplying material for the AGN.

\begin{figure*}
\centering
\includegraphics[width=0.75\textwidth, angle=0]{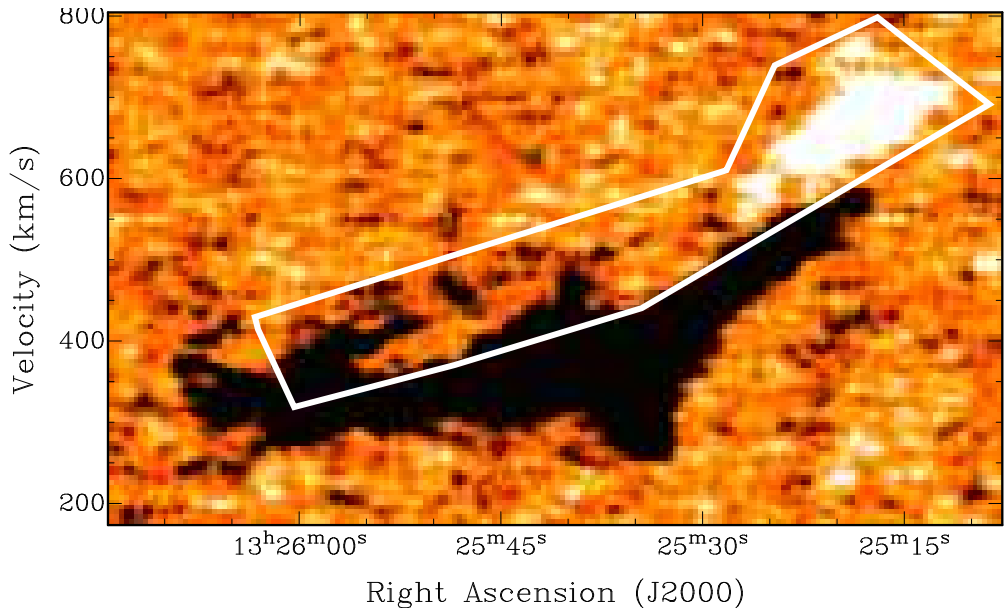}
\caption{Position-velocity plot obtained along a slit in $PA=90^{\circ}$ as illustrated in Fig.~\ref{mom0.spec}. The data is integrated over 24\arcsec ~in Declination to better highlight the faint outer \HI arm structure in emission (black, inside the white box) which connects with the absorption (white) on the western side. Note that the greyscale is inverted (i.e. absorption white, emission black) compared to Fig.~\ref{CO.HI}, \ref{absJet} and \ref{pvplots.models} to better highlight the faint emission.}
\label{pv.plot2}
\end{figure*}

Interestingly, the radial motions of the outer regions are such that the projected velocities are {\sl redshifted} compared to the regular orbits (see below and Fig.~\ref{pv.plot2}). This means that the blueshifted absorption discovered in Paper I (see also Sect.~3.2) cannot be caused by out-moving gas at large radius projected onto the centre. Therefore, the interpretation of the blueshifted absorption, together with the already known redshifted nuclear absorption, as evidence for a regular inner disk, can still hold.

Previous studies have already shown that the kinematics of the gas in the dust lane is dominated by rotation (see $pv$-diagram in Fig.~2 of Paper~I). Although there is a fair degree of symmetry in the \HI distribution, significant deviations from symmetry are  present, not only in morphology, but also in kinematics. As can be seen from Fig.~\ref{mom0.spec}, the famous optical dust lane has a clear counterpart in \Hi . Below the dust lane, south-west of the nucleus, a similar amount of \HI is present as  in the dust lane itself. Previous observations \citep[e.g.][]{schiminovich94} had already shown the presence of this \Hi , but it is better visible in our data. Most of the \HI south-west of the centre can be explained in a simple way. In the optical, the dust lane is only visible where the gas and dust disk is in front of the galaxy, i.e.\ the north-eastern side. South-west of the nucleus, the disk is behind most of the stellar body and it will be much less visible in absorption. However, the \HI of the disk is seen in emission and therefore both the back and the front side of the disk are visible equally well.

The kinematics of the \HI also shows that the outer gas disk
south of the centre is not perfectly regular. The position-velocity plot given
in Fig.~\ref{pv.plot2} shows the kinematics of part of the south-western disk. Since the
position-velocity plot was taken far from the nucleus, both the emission and
absorption should reflect the kinematics of the same outer \HI disk. The only
difference between absorption and emission should be that the absorption  comes
from that part of the south-western disk that happens to be in front of the radio continuum lobe, but the kinematics of the absorption should smoothly fit that of the
emission. However, the absorption is offset in velocity with respect to the emission, something that is not expected for a regular rotating disk with circular orbits. In addition, the absorption structure has a faint continuation in emission, see the box in Fig.~\ref{pv.plot2}. 

The presence of this gas is interesting for various reasons. The part in emission represents a spatially and kinematically continuous structure that appears to be, at least partly, connected to the large-scale disk.  
The gas originating the emission and the absorption (inside the box in Fig.~\ref{pv.plot2}) is likely to be one continuous structure, representing front and back of one continuous structure (called ``outer arm''). The observed {\sl broad} profiles suggest that radial motions must be affecting the gas in addition to the rotation. Further intriguing is the fact that we do not detect similar structures on the opposite (northern) side of the dust lane. This already points to an asymmetry at large radius in the distribution of gas that is clearly a complication in the modelling of the data (see Sect.~4).

%__________________________________________________________________ %
%_____________________________________________________________
%                                             Two column Table 
%_____________________________________________________________
%
\begin{table}
\caption{Absorption-line properties.}
\label{absorp.table}
\centering          
\begin{tabular}{c c c l l l }
\hline\hline       
Component & V   & $\Delta v$  & $\tau$  & $N_{\HI}$/$T_{\rm{S}}$  \\
                &  [km/s]  &  [km/s]  &  &  \\
                &  (1) &   (2) & (3) & (4) \\
\hline                    
Nucleus & 550 & 18 & 0.83 & $2.74\times 10^{19}$ \\
Nucleus & 574 & 23 & 0.62 & $2.61\times 10^{19}$ \\
Nucleus & 592 & 20 & 0.26 & $9.52\times 10^{18}$ \\
Nucleus (shallow) & * & 400 & * & *  \\
Jet & 550 & 19 & 0.64 & $2.23\times 10^{19}$ \\
Jet & 544 & 25 & 0.64 & $2.93\times 10^{19}$ \\
Cloud & 454 & 13 & 0.07 & $1.72\times 10^{18}$  \\
\hline                  
\end{tabular}
\caption*{(1) Velocity of the peak. (2) FWHM (3) Peak optical depth. (4) \HI column density [cm$^{-2}$~K$^{-1}$].}
\end{table}

%__________________________________________________________________
%
\begin{figure}
\centering
\includegraphics[width=0.34\textwidth, angle=270]{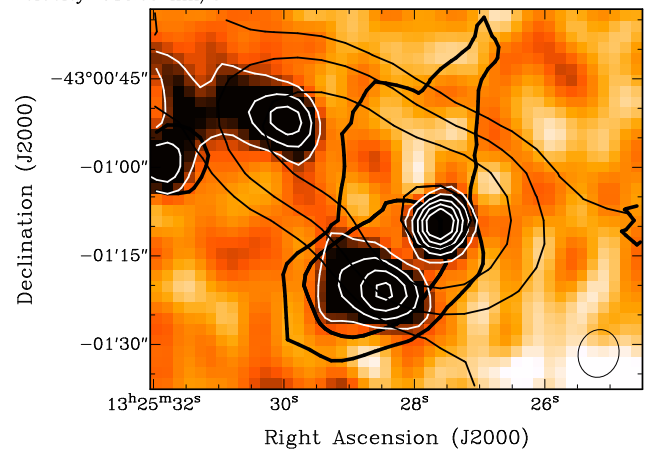}
\caption{Velocity channel of the high-resolution cube (greyscale). The white contours show the \HI absorption (-15, -12, -9, -6 and $-3\sigma$). The thick black contours give the CO emission \citet{liszt01}. In thin black contours (0.5, 1.0, 2.0 and 4.0~Jy~beam$^{-1}$) the low-resolution continuum image is shown. The beam of the \HI cube is indicated in the lower right corner.}
\label{CO.HI}
\end{figure}
%__________________________________________________________________
%

\subsection{\HI in absorption}

The ATCA observations reveal new features also for the \HI absorption. Figure~\ref{m0.optical} shows the total intensity of the \HI in absorption as well as the low-resolution continuum as obtained from the line-free channels. The radio continuum shows the inner lobes and the inner jet, in agreement with previous observations \citep[e.g.][]{clarke92,burns83}.
The \HI absorption is detected against the southern radio lobe, the northern jet and the unresolved nucleus, see Fig.~\ref{mom0.spec} for more details. In addition, we detect absorption up to $\sim 20$\arcsec ~(corresponding to $\sim 3$ beams) east of the core, in the direction perpendicular to the jet (Fig.~\ref{CO.HI}), indicating the presence of background radio continuum extended in that direction. There is a close correspondence between this absorption and the CO emission as detected by \citet{liszt01} (see Fig.~\ref{CO.HI}).

\begin{figure}
\centering
\includegraphics[width=0.47\textwidth]{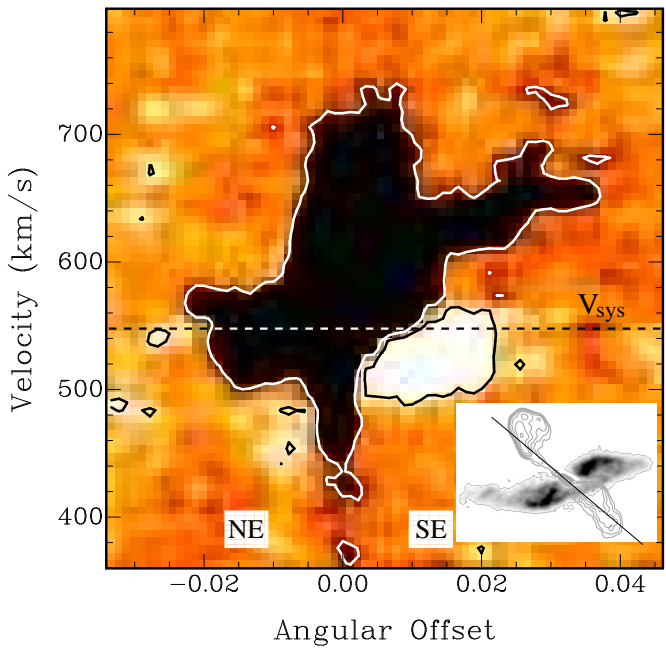}
\caption{Position-velocity plot obtained along the radio jet, as illustrated in the inset in the lower right corner. Contour levels: -3 (white) and $3\sigma$ (black). The systemic velocity is indicated by the dashed line.}
\label{absJet}
\end{figure}

Figure~\ref{absJet} shows a position-velocity plot along the jet PA. This illustrates how complex the absorption is  and, in particular, that there is redshifted absorption everywhere. The outer filament at large radius is redshifted and partly in front of nucleus (see Sect.~3.1 and Fig.~\ref{pv.plot2}). It is therefore difficult to disentangle whether at least part of the redshifted absorption is at all associated to infalling gas. No equivalent,  large-scale blueshifted gas seems to be present. 

%__________________________________________________________________
%
%
%__________________________________________________________________
%

As shown in Fig.~\ref{mom0.spec}, the absorption against the northern jet and the beginning southern lobe has a typical column density (assuming $T_{\rm{spin}}=100$~K) of the order of $10^{21}$~cm$^{-2}$ which is of the same order as the column densities of the neighbourhing emission. Because both seem to be connected to the large-scale disk (Sect.~4) this suggests that 100~K is a reasonable choice for the spin temperature in this region. However, the south-western parts of the absorption have more than an order of magnitude lower column densities - i.e. $\le 10^{20}$~atoms~cm$^{-2}$ (see Fig.~\ref{mom0.spec}) - and hence these column densities are below the detection limit ($3\sigma$ level per channel) of the \HI emission.

While the extended \HI absorption against the northern jet and the southern lobe is in agreement with previous observations, most exciting is the result that the absorption against the unresolved nucleus is much broader than previously known (Fig.~\ref{mom0.spec}). In particular, the blueshifted part of the absorption was undetected in previous studies. A complete analysis and discussion of the origin and nature of this absorption against the nucleus has been presented in Paper~I. 
While the shallow part of the nuclear absorption is new, the deep components close to the systemic velocity are in agreement with \citet{vdh83}. Peak velocities, width of the profiles, optical depths and column densities are summarized in Tab.~\ref{absorp.table}.

For the nuclear components of the \HI absorption, the value of the spin temperature is much more uncertain as the gas could be located closer to the AGN (see Sect.~4 and 5) and therefore $T_{\rm{spin}} = 100$~K could  underestimate the real value. According to \citet{bahcall69} and \citet{maloney96} the spin temperature can easily reach a few times $10^3$~K for gas close to an AGN. Assuming a spin temperature of a few 1000~K, we estimate column densities of $\sim 10^{22}-10^{23}$~cm$^{-2}$, similar to what is found in other radio galaxies  \citep[e.g.][]{peck01, momjian02, momjian03, morganti04}.
We can compare the \HI column density of Cen~A with the value derived from X-ray observations. \citet{evans04} have estimated a column density for the nucleus (from Chandra and XMM-Newton data) of $10^{23}$~atoms~cm$^{-2}$ that is comparable to our value if we assume $T_{\rm{spin}}\approx 3000$~K. It is worth noting, that the column density derived for Cen~A from the X-ray is relatively high compared to other radio galaxies of similar radio power \citep[see e.g.][]{balmaverde08}.

%__________________________________________________________________
%
\begin{figure*}
\centering
\includegraphics[width=0.85\textwidth]{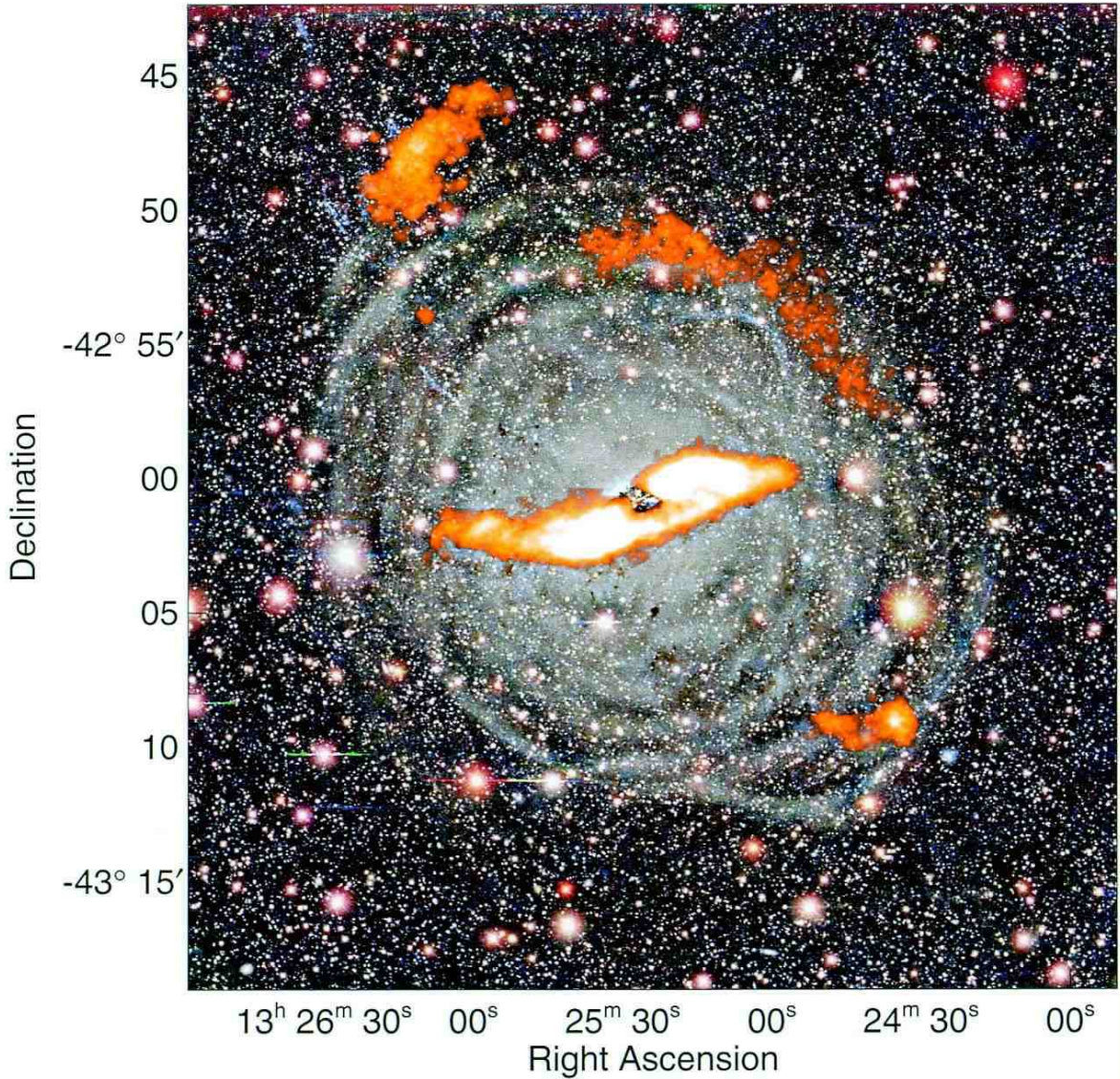}
\caption{\HI emission obtained from the ATCA observations (greyscale; orange in online version) superimposed to an optical BVR image obtained after unsharp masking and adaptive histogram equalisation by \citet{peng02}.}
\label{m0.emission}
\end{figure*}
%
%__________________________________________________________________
%

\subsection{Two newly discovered \HI clouds in the environment of Cen~A}

\HI at large distances from the dust lane was detected and studied by \citet{schiminovich94}. These large-scale structures are also detected in our data as illustrated in Fig.~\ref{m0.emission} which shows an overlay of the \HI emission on an optical image obtained by \citet{peng02}. The optical image emphasises the low-contrast features and in particular the complex and extended set of shells and the faint dust extensions. However, our observations include only one pointing on the centre of Cen~A and therefore these structures are not imaged as well as in the work of Schiminovich. We therefore consider the study of this large-scale \HI structure beyond the scope of this paper.

However, it is interesting to note that two additional unresolved clouds are detected in our observations, far away from the \HI disk (see Fig.~\ref{m0.optical}). The first cloud is detected in absorption against the NE radio-lobe (see also Fig.~\ref{mom0.spec}) at a distance of 4\farcmin7 (about 5.2~kpc) from the nucleus (PA~=~50\fdegr 5) and has a velocity width (FWHM) of only $\Delta v=19$~\kms . The velocity of this cloud (454~\kms ) is below the systemic velocity of Cen~A. The optical depth is $\tau = 0.05$ corresponding to a column density of about $10^{20}$~atoms~cm$^{-2}$, resulting in an upper detection limit of $10^5$~\Msun ~(assuming a spin temperature of 100~K and a filling factor of unity).
The second cloud is located 11~kpc from the nucleus at a position angle of about 48\fdegr 5 w.r.t. the nucleus and is detected in emission at $v=418$~\kms. The cloud is only visible in the low resolution cube, which indicates a diffuse nature. Also this cloud has a small velocity width ($\Delta v=24$~\kms ) and an \HI mass of about $8\times 10^5$~\Msun . Both clouds are located along the jet axis where also ionised gas has been detected \citep[e.g.][]{dufour78,morganti91}.

%__________________________________________________________________
%
\begin{figure}
\centering
\includegraphics[width=0.43\textwidth]{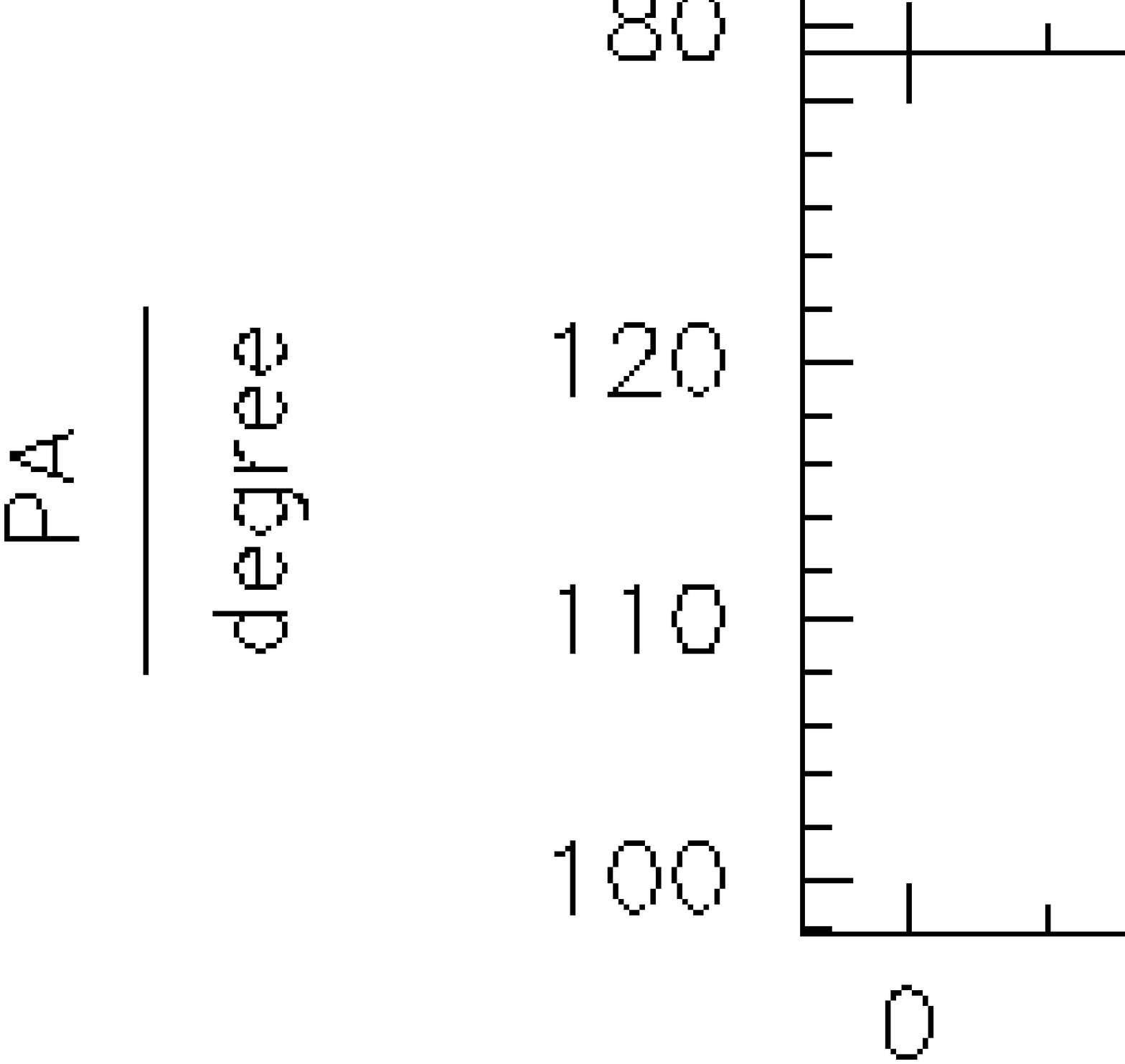}
\caption{Parameters of the \HI model described in Sect.~4.}
\label{model.parameters}
\end{figure}
%__________________________________________________________________

%_____________________________________________________________
%
\begin{figure*}
 \centering
\includegraphics[width=0.9\textwidth]{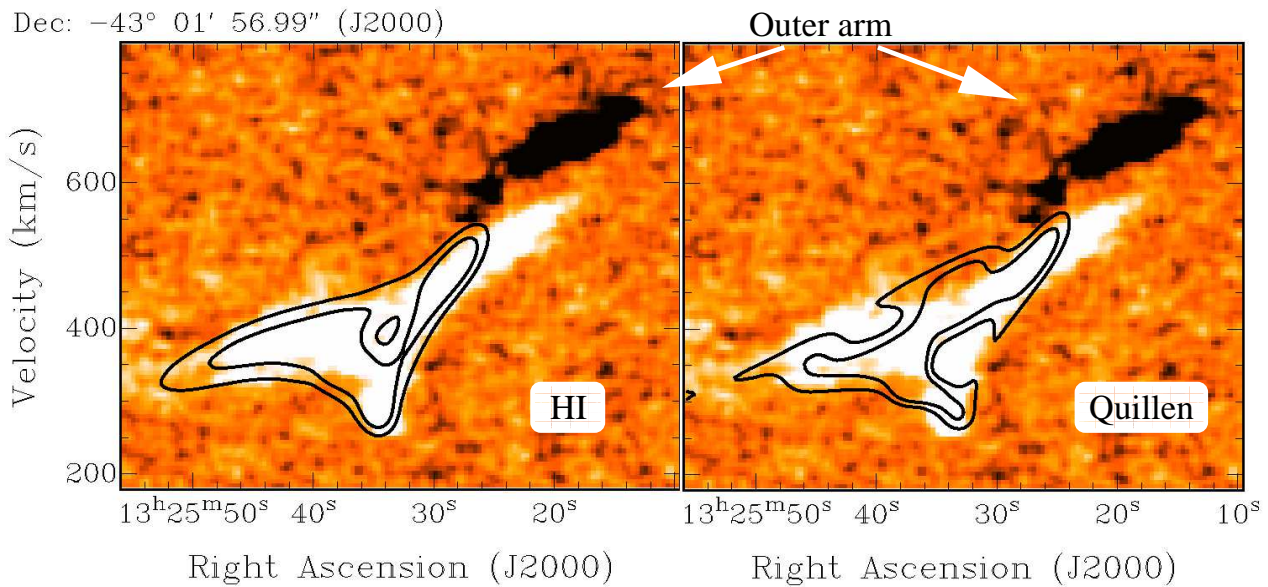} 
  \caption{Position-velocity plot along the $PA$ indicated in Fig.~\ref{mom0.spec}.}
  \label{pvplots.models}
\end{figure*}
%_____________________________________________________________
%

\section{The H~{\small I} model}

The modelling of the \HI has at least two main goals. It allows to investigate the formation and evolution  of Cen~A (e.g. its merger history) through the study of the morphology and  kinematics of the gas over the whole disk. The molecular and ionised gas disks are not as extended as the \Hi . The second reason is to pin down  the location of the gas responsible for the nuclear absorption. Is the absorbing gas located in the vicinity of the nucleus or at large distance projected in front of the core? The answer to this question is important for the understanding of the AGN activity (see Sect.~1 and Paper~I).

A number of modelling studies of the large-scale gas disk of Cen~A have been done in the past \citep[see][for a review]{morganti10}, but previous \HI emission-line studies \citep{vangorkom90} did not allow such a detailed investigation as presented here. \citet{bland87} and \citet{nicholson92} used the velocity field derived from H$\alpha$ observations to obtain models describing the warped dust-lane structure with a set of tilted-rings. The largest radius fitted corresponds to about 245\arcsec ~($\sim 4.5$~kpc see also Fig.~\ref{incl}).
\citet{quillen92} obtained from CO observations a first warped disk model that matches the optical appearance of the dust lane and which is also consistent with the \HI rotation curve derived by \citet{vangorkom90}. In the most recent model of  \citet{quillen06A} derived from new mid-infrared observations, the position angle is quite similar --- at large radii ---  to the one of \citet{nicholson92} although the inclination is closer to edge-on for $r>140$\arcsec ~(see also Fig.~\ref{incl}).

\citet{sparke96} developed a number of dynamical models to explain the geometry of the disk as that of a near-polar structure precessing around the symmetry axis of an approximately oblate galaxy potential. These models compare well with the orientation of the model by \citet{nicholson92}.
According to all these models, the gas disk is strongly warped and crosses the line-of-sight several times at different radii. With this complex geometry, it is impossible to unambiguously assign velocities to a position on the sky and, therefore, to derive a reliable tilted-ring model \citep{rogstad72} and a rotation curve based on the velocity field only. Thus, a construction of model cubes is needed for a proper comparison with the data.

%_____________________________________________________________
%
\begin{figure*}
 \centering
 \includegraphics[height=8cm]{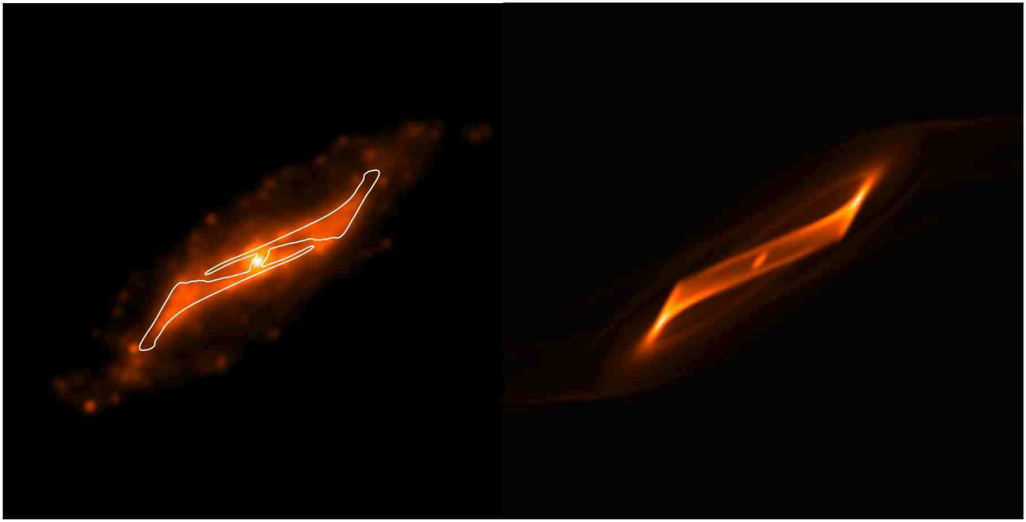}
  \caption{Left panel: Spitzer image \citep{quillen06A} over plotted with intensity contours of the high resolution \HI model. Right panel: Total intensity map of the \HI model at the resolution of the Spitzer data.}
  \label{CompSpitzer}
\end{figure*}
%_____________________________________________________________
%

To achieve this we have created full 3D model cubes using GASMOD, a phyton code to calculate the tilted-ring models (kindly provided by Paolo Serra).
The initial parameters of inclination and position angle for the large-scale disk were taken from \citet{quillen06A}. Because  these parameters do not extend to the entire range of radii covered by the \Hi , we have extrapolated them to larger radii.
A flat rotation curve, that rises steeply inside our first ring has been assumed.
The best values for the coordinates of the centre and for the systemic velocity resulted in agreement with the value from literature values. For the systemic velocity the value of $v_{\rm{sys}}=546$~\kms ~was used, similar to the value of \citet[][$v_{\rm{sys}}=542$~\kms ]{vangorkom90}. All parameters (surface brightness, rotation velocity, inclination, position angle) were then independently varied from their initial values for every single ring. 

The model cubes were  compared with the \HI data cube to make sure that the model would describe the kinematics of the observed gas. The changes to the model were made based on a visual inspection of model cube and data. We also produced from the model the distribution of the total intensity of the \HI that was then compared with the Spitzer image in order to make sure that the model would also reproduce the overall characteristics of this image.
In order to perform this comparison, models were produced with the spatial resolution of the Spitzer data (see below).

The starting parameters clearly did not provide a satisfactory description of the data.  The final value of the parameters as function of the radius are shown in Fig.~\ref{model.parameters}. The rotation curve is essentially flat and midly decreasing with radius, a phenomenon also observed in other early-type galaxies \citep[see e.g.][]{noordermeer07}. Figure \ref{pvplots.models} shows a comparison between a slice of the data and the starting and final model. The slice is close to what is shown in Fig. \ref{pv.plot2} and, therefore, it includes some of the unsettled gas described in Sect.~3.1 that we do not expect to see described by the model. However, the region of strong \HI emission is clearly better reproduced by our \HI model (figure on the left) compared to the model of \citet{quillen06A}. In addition, our \HI model also describes the dust morphology (Spitzer image) as shown in Fig.~\ref{CompSpitzer}.

Thus, the model (Fig.~\ref{model.parameters}) describes most of the \HI morphology and kinematics confirming the regular rotation of the \HI structure for $r<6$~kpc. The orientation parameters (position angle and inclination) confirm the heavily warped structure of the disk. In addition, the model shows that in \HI a gap in the distribution of the gas is present between 10\arcsec ~and $\sim 45$\arcsec ~(from 0.2 to 0.8 kpc), confirming what was already found by other studies  \citep[e.g.][]{nicholson92,quillen92,quillen06A,espada09} for the dust, molecular and ionised gas (e.g. H$\alpha$ and  CO).

Our best model uses parameters different from what was found in previous modelling. Figure~\ref{incl} shows a comparison between the parameters of our best model and the parameters from \citet{quillen06A}. The main noticeable difference is in the inclination that appears to stay closer to edge-on for $r > 90$\arcsec ~in our model. 
Remarkable is the similarity of our parameters with the independent results from the study of the stellar light emerging from the kiloparsec-scale, ring-like structure  detected using the VLT \citep{kainulainen09}.

Our model description of the data is solely based on circular rotation and does not require a bar structure \citep[previously claimed to be present by][]{mirabel99}, in agreement with the results from the Spitzer images of \citet{keene04} and \citet{quillen06A}.

Recently, \citet{espada09} have modelled high-resolution CO data obtained with the SMA for the central kpc of Cen ~A and invoked the presence of a weak bar  in these central regions.  Interestingly, while their model seems to very poorly describe the jump in velocity observed in the CO velocity field at about 15" from the centre (see their Fig.~4 compared with their Fig.~11), our model naturally explains this as only due to the geometry of the disk. The presence of a small-scale bar is also ruled out by the near-IR data presented by \citet{neumayer07}.

\section{Discussion}

The new data presented here have allowed to image in detail the \HI emission and absorption of the warped disk of Cen~A and to identify new features in the gas. In this section we discuss the implications of the \HI results for the formation and evolution (e.g. merger history) of Cen~A and the different phases of AGN activity.

%__________________________________________________________________
%
\begin{figure}
\centering
\includegraphics[height=8cm, angle=270]{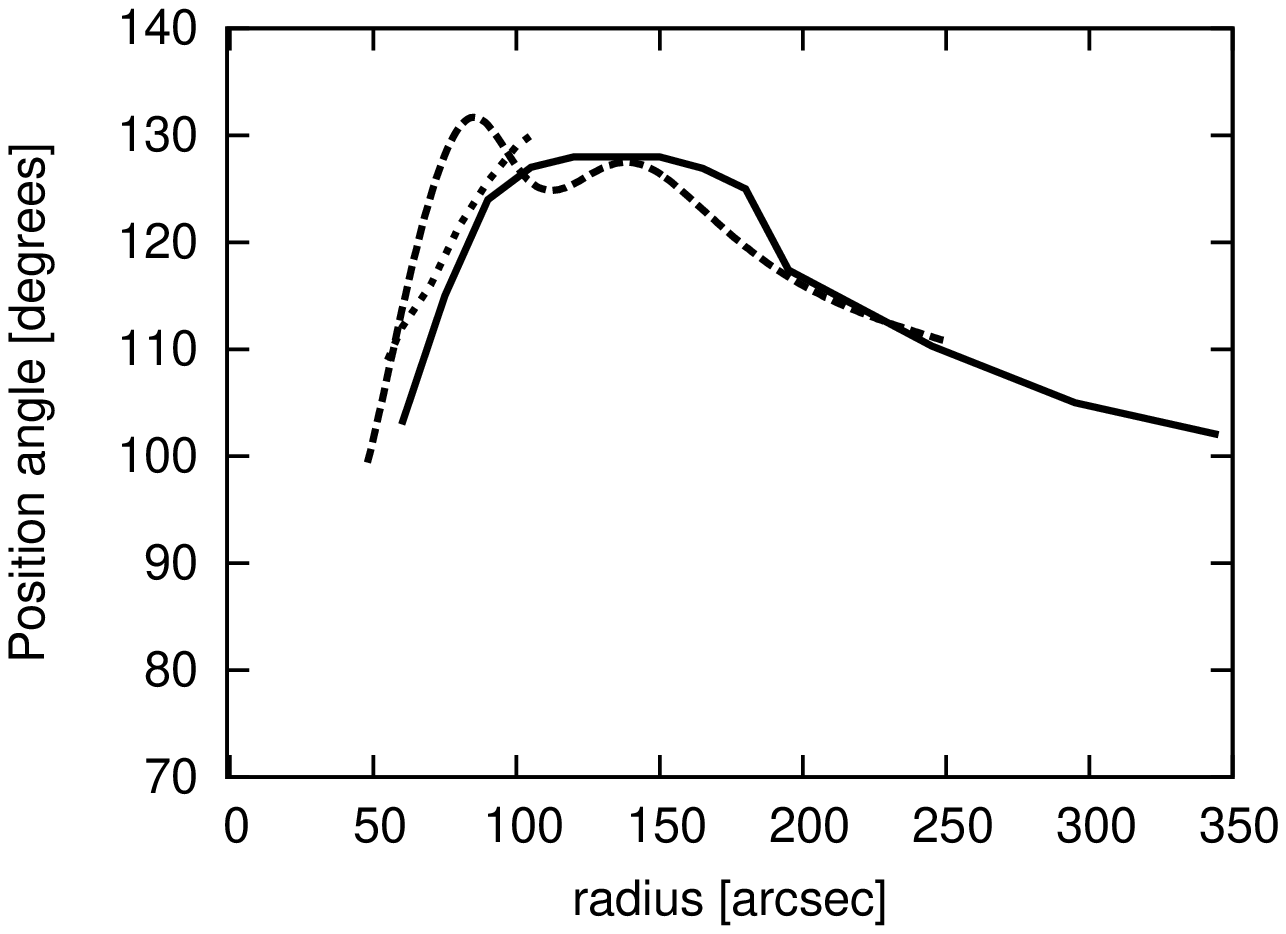}
\includegraphics[height=8cm, angle=270]{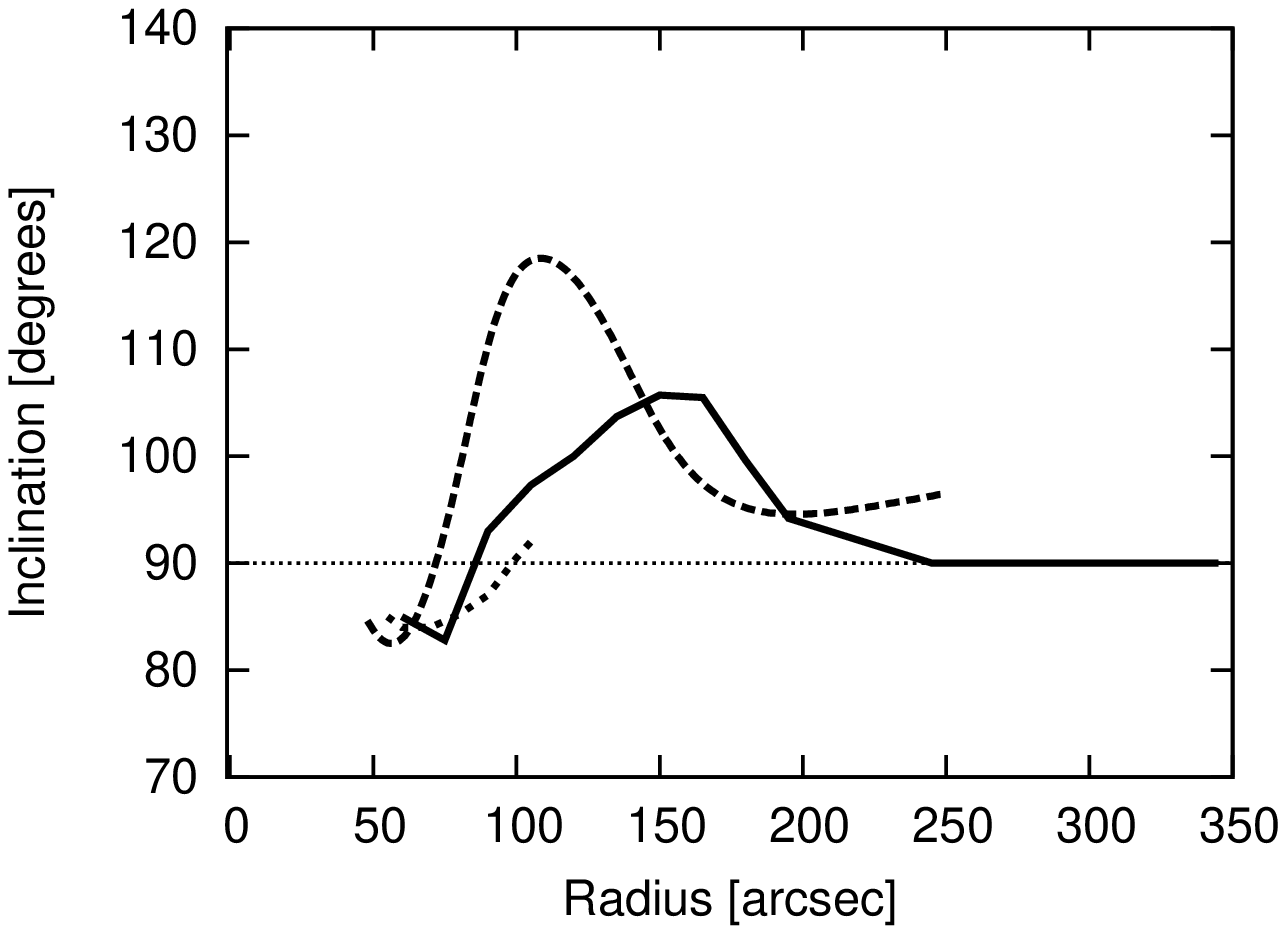}
\caption{Position angle (top panel), inclination (bottom panel) of our \HI model (solid line) superimposed to the values from the \citet{quillen06A} model (long dashed line, similar to those of \citet{nicholson92}. In addition, also the values from the stellar modelling of \citet{kainulainen09} are plotted (dotted line).} 
\label{incl}
\end{figure}
%__________________________________________________________________
%

\subsection{Modelling implications: The large-scale \HI distribution} 

Our data confirm that the structure and kinematics of the \HI in Cen~A is dominated by rotation, down to the central regions, and most of the \HI (but clearly not all) is  confined to a fairly regular, warped disk (for $r<6$~kpc). 

In addition to this, we detect outer unsettled gas that appears to be located in an arm-like structure. Therefore, some of the gas in emission and absorption is observed below the dust-lane and it is not reproduced by our modelling. This arm shows a smooth velocity gradient - similar for absorption and emission but with an offset in velocity  between the two. Because of the regular velocity characteristics of this structure, it is likely that it will become part of the outer part of the rotating disk in the near future. 

This type of structure is expected as a result of a merger or accretion as shown in many numerical simulations \citep[e.g.][]{hibbard96,barnes02}. Thus, the \HI further support the hypothesis of Cen~A being the result of a relatively recent merger of an elliptical galaxy with a smaller gas-rich (SMC-like?) disk galaxy \citep[e.g.][see also Sect.~1]{baade54,quillen93}. 

Using the results of our modelling, we can derive a rough estimate of how long ago the merger must have taken place to allow a large fraction of the \HI to settle down in the warped disk we observe.
We can assume that the gas located at larger radii than we can model (i.e. $r>350$\arcsec ) has performed at least one or two revolutions since the merger took place giving a rough estimate resulting in $1.6-3.2\times 10^8$~yr. This time estimate is in agreement, inside the uncertainties,  with \citet[][$2\times 10^8$~yr]{quillen93} and \citet[][$3\times10^8$~yr]{peng02}, while it is  shorter than what derived by \citet{sparke96} - $7.5\times 10^8$~yr -  that would  correspond to almost five revolutions. Such a long timescale would be sufficient to relax most of the gas in the disk, in contradiction with our \HI results.

The merger hypothesis does find further support from the partial ring structure discovered by \citet[][see also Sect.~1]{schiminovich94}. In that respect the existence of additional, low mass ($<10^6$~\Msun ) \HI clouds is not surprising. The velocities of the two newly discovered \HI clouds (see Sect.~3.3) are in rough agreement with the spatial velocity distribution of the ``outer ring'' structure. 
Clouds and filaments of ionised gas have been found in this region along the jet direction \citep[e.g.][]{dufour78,morganti91}. For the origin of these structures, two hypothesis have been put forward. In addition to the merger origin, an alternative scenario was the cooling condensations in a tenuous, thermally unstable hot component of the interstellar medium gas. This scenario was somewhat favoured to explain the filaments of ionised gas, as it would explain the highly turbulent velocity structure observed there \citep[see also][]{oosterloo05}. However, such high velocity gradients are not seen in the \HI profile of the newly detected clouds which could mean that the clouds were not hit by the jet. Whether the newly discovered \HI clouds are a cooling condensation or are more likely the left over of the interaction/merger discussed above, remains unclear.

As a final remark, we would like to note that also the distribution of the \HI confirms what was found by other studies of the  molecular gas and dust  \citep{nicholson92,quillen06A}, i.e. the lack of gas in region between  10\arcsec ~and  45\arcsec ~(0.2~kpc to 0.8~kpc) while on the smallest scales circumnuclear disks are found (Paper~I and ref. therein). 
The peak in the surface brightness around 150\arcsec ~from the centre suggests a ring-like structure in the distribution of \HI in Cen~A (see Fig.~\ref{model.parameters}). \HI ring-like structures are often detected around early-type galaxies and there are a number of possible scenarios for the formation of such a structure (see e.g. Donovan et al. 2009 for a summary and references). A ring morphology can be the result of the presence of a bar structure but we found no evidence for such a structure in Cen~A and indeed many other galaxies with gas rings do not show bars. In some cases, "burning out" of the gas in the central regions of the galaxies has been proposed, but the famous example of IC~2006 \citep{schweizer89} shows that also gas accretion can form these structures.

\subsection{Merger and different phases of radio activity}

The kinematics of the \HI in Cen~A is dominated by rotation down to the nuclear regions. The discovery that the nuclear absorption is also partly blueshifted, (see Paper~I) has questioned the interpretation that \HI against the nucleus represent evidence for the {\sl direct} fuelling of the AGN. Furthermore, our analysis presented here shows that the nuclear blueshifted absorption is not due to gas at large radius seen in projection against the core. The nuclear absorption is likely originating from a nuclear disk that represents the \HI counterpart of the structure seen from the molecular gas (see Paper~I for a detailed discussion). This nuclear disk is part of the overall warped disk structure. Thus, our results confirm what suggested in Paper~I that the nuclear \HI gas does not constitute direct evidence of gas infall into the AGN. 

However, it is worth mentioning that radial motions (in addition to rotation) on the scale of tens of pc have been detected in the high resolution, near-IR data obtained by \citet{neumayer07}. Surprisingly  they found that for higher excitation lines [SiVI] and [FeII] the velocity pattern is increasingly dominated by a non-rotating component, elongated along the radio jet. Interestingly, these non-rotational motions were detected {\sl along} the jet with redshifted velocities (compared to the systemic) seen on the main-jet side  and blueshifted on the counter-jet side. These motions (stronger in [SiVI]) can be explained as backflow of gas accelerated by the plasma jet.  This gas can perhaps be involved in the fuelling of the AGN. Whether a similar situation is also occurring in the \HI gas can only be investigated with deep VLBI observations.

In order to investigate in more detail whether a link can be found between the merger/accretion event and the nuclear activity, it is important to compare the timescales of these events. Cen~A shows a complex radio structure with lobes on different scales, from a few kpc of the inner lobe (see Fig.~\ref{m0.optical}) to tens of kpc for the middle lobe and up to several hundred kpc for the outer lobe (Sect.~1).  These three radio structures could be either the result of precession \citep[][see also Sect.~1]{haynes83} or the nuclear activity in Cen~A has been recurrent. However, the large-scale structure is currently experiencing a particle injection event, as recent spectral-index studies suggest \citep[][and refs. therein]{hardcastle09}.

The ages of these lobes are not easy to estimate. According to \citet[]{saxton01} the age of the northern {\sl middle} lobe can be estimated to be about $1.4\times 10^8$~yr assuming this lobe has been created by an old episode of jet activity that has been {\sl interrupted} by the disruption of the jet by the merger. Since then, this structure has been rising buoyantly away from the nucleus. If this is the case, the merger that has brought the \HI would actually be the cause of the disruption of the old radio activity. The timescale of the merger derived from the \HI could support this idea. The timescale of the inner radio lobe, $\sim 5-10\times 10^6$~yr \citep[][and refs. therein]{croston09}, would indicate that it takes quite some time for the system to recover from this disrupting event and for the gas to settle again in a steady accretion flow. A (possible) time-delay between the merger and the onset of the AGN has also been detected in a few other nearby radio galaxies \citep[e.g.][]{tadhunter96,emonts06,davies07,wild10}.

It is worth noting that if the scenario proposed by \citet[]{saxton01} is correct, the delay between the disruptive merger and the onset of a new phase of activity could be related to the time required by the X-ray halo to build up again \citep[see][]{sansom00}. 

On the other hand, recent spectral-index studies have suggested different timescales for the large-scale radio structure. High frequency studies show a spectral break for the southern lobe that would suggest a life time of  $3 \times 10^7$ yrs \citep[see][]{hardcastle09}. Furthermore, no spectral break is observed in the northern lobe (both the giant and the middle lobe), suggesting  that they have very recently undergone some particle injection event \citep[][and refs. therein]{alvarez00,hardcastle09}.  Thus, if the timescales derived from the spectral index are correct, it would indicate that that the {\sl overall} radio emission starts relatively late compared to the merger. 

In both cases, the comparison of the timescale derived from the \HI and the radio structure, together with the relatively regularity of the \HI down to the sub-kpc regions, do not suggest a one-to-one correspondence between the merger and the phase of radio activity. In particular, if the timescales derived from spectral index are correct, {\sl more than one radio burst can be produced while the merger is settling in}. This would indicate that, while the merger could be responsible for bringing gas in the vicinity of the nucleus, the accretion mechanism that fuels the AGN is likely not related to the merger. 

This is in agreement with the idea that in radio galaxies of relatively low radio power (i.e. FR-I type) the fuelling of the black hole proceeds directly from the hot phase of the interstellar medium in a manner analogous to the Bondi accretion. As suggested by a number of authors \citep[see e.g.][]{allen06,hardcastle07} the hot, X-ray emitting phase of the intergalactic medium (IGM) is sufficient to power the jets of several nearby, low-power radio galaxies. In the case of Cen~A, a mass accretion rate $\dot M_{Bondi} = 6.4 \times 10^{-4}$\Msun ~yr$^{-1}$ and a Bondi efficiency of $\sim 0.2$ \% has been derived from Chandra observations \citep{evans04}. However, it would decreases if the new value of the BH mass derived by \citet{neumayer07} and \citet{cappellari09} is  used.
These values are consistent with what is found for other FR-I radio galaxies \citep{balmaverde08} while the efficiency appears to be lower than the "canonical" 0.1 for the high efficiency optically thick accretion disks.  On the other hand, and to complicate the picture, it is known \citep[from the Fe K$\alpha$ line and the large column density measured in the X-ray towards the nucleus][]{evans04} that Cen A has large quantities of cold gas in a molecular torus at about 1pc from the black hole, more typical of FR-II galaxies. 

%__________________________________________________________________
%

\section{Summary}

In this paper we have reported ATCA \HI synthesis observations which have higher spatial and velocity resolution than previous studies of atomic hydrogen in Cen~A. Our study concentrates mainly on the distribution and kinematics of the gas associated with the dust lane of this galaxy. These data show that the kinematics of the \HI in Cen~A is dominated by rotation down to the nuclear regions.

The warped morphology and kinematics of the \HI can be described with a tilted-ring model with gas on circular orbits. Regular kinematics of the gas is observed for $r<6$~kpc. We find no need for non-circular motions between 1-6~kpc.
There is a lack of \HI between $0.1-1$~kpc, confirming results from observations of CO,  ionised gas and dust.
The parameters of the model (in particular the inclination) are different from what derived in previous models \citep{quillen06A,nicholson92} but nicely consistent with those from the study of the central stellar ring of \citet{kainulainen09}.  
Outside 6~kpc radius, asymmetries and arm structures are observed as well as radial motions. 

The \HI fraction in Cen~A ($M_{\HI}/ L_B = 0.01$) is relatively low for an early-type galaxy, indicating that the \HI disk might well be the result of accretion of a small (SMC-like?) galaxy. Based on the regular kinematics we suggest that this accretion occured about $1.6 - 3.2 \times10^8$~yr ago, in agreement with previous estimates. This is too old to {\sl directly} trigger the current phase of AGN activity (the age is a few $\times 10^6$ yr for the inner radio lobes), but might have interrupted the previous episode of AGN activity.
 
We find extensive \HI absorption, against nucleus, jet and lobes. Part of the absorption (in particular with velocities close to $v_{\rm{sys}}$) is part of the regular rotating disk, while some of the absorption against the southern lobe is part of an outer arm structure. A large fraction of the {\sl nuclear} absorption is not produced by outer gas confirming our hypothesis from Paper~I. This is in particular the case for the newly discovered blueshifted absorption. This absorption is, at least partly, due to a circumnuclear disk at $r < 100$~pc. Whether part of the (redshifted) gas is infalling and fuelling the AGN is difficult to establish with our data. The results of \citet{neumayer07} --- who detected radial motions along the jet axis in near-IR observations --- suggest that it could be worth exploring with deep VLBI data whether such a component also exists in \Hi .

All in all these results suggest that in galaxies with an AGN where a merger/accretion event has clearly occurred in the recent past, the connection between these two events (merger and activity) is far from clear. 
In the case of Cen~A, the AGN may instead be fuelled via a low efficiency/rate accretion, as suggested for other FR~I galaxies of similar radio power. The merger may actually have played the opposite role of temporarily disrupting the radio jet and producing the northern middle radio lobe. The delay in creating the much younger inner radio lobe, may be connected to the time that it takes the gas to settle again in a steady accretion flow.

%__________________________________________________________________

\begin{acknowledgements}

We would like to thank Paolo Serra for kindly providing his code to calculate the tilted-ring models, GASMOD. His code can be found under: http://www.astron.nl/$\sim$serra/ We thank Alice Quillen for providing her model in digital form and the Spitzer image in fits format. This work is based on observations with the Australia Telescope Compact Array (ATCA), which is operated by the CSIRO Australia Telescope National Facility. This research was supported by the EU Framework 6 Marie Curie Early Stage Training programme under contract number MEST-CT-2005-19669 "ESTRELA".

\end{acknowledgements}

\end{document}